\documentclass[twocolumn,showpacs,superscriptaddress,preprintnumbers,prd]{revtex4}
\usepackage{graphicx,amsmath,dcolumn,bm}
\begin{document}
\preprint{KEK-TH-1269}
\title{\Large \bf Projections of structure functions in
                  a spin-one hadron}
\author{T.-Y. Kimura}
\affiliation{Department of Particle and Nuclear Studies,
             Graduate University for Advanced Studies \\
             1-1, Ooho, Tsukuba, Ibaraki, 305-0801, Japan}      
\author{S. Kumano}
\affiliation{Institute of Particle and Nuclear Studies,
          High Energy Accelerator Research Organization (KEK) \\
          1-1, Ooho, Tsukuba, Ibaraki, 305-0801, Japan}
\affiliation{Department of Particle and Nuclear Studies,
             Graduate University for Advanced Studies \\
           1-1, Ooho, Tsukuba, Ibaraki, 305-0801, Japan}     
\date{September 29, 2008}
\begin{abstract}
There exist new polarized structure functions in a spin-one hadron.
In deep inelastic electron scattering from a spin-one hadron, there are
eight structure functions 
$F_1$, $F_2$, $g_1$, $g_2$, $b_1$, $b_2$, $b_3$, and $b_4$.
We derive projections to these eight functions from the hadron tensor
$W^{\mu\nu}$ by combinations of the hadron momentum and its polarization
vectors.
\end{abstract}

\maketitle

\vspace{-0.2cm}
\section{Introduction}
\vspace{-0.2cm}

High-energy spin physics became an important topic since the discovery
of the European Muon Collaboration (EMC) spin effect \cite{emc88}. 
It indicated that almost none of the nucleon spin is carried by quarks,
whereas the total amount should be carried in a naive quark model.
Since the EMC finding, many experiments have been done for the nucleon
spin. We now have a rough idea about each quark-spin contribution to 
the nucleon spin \cite{ppdfs}. However, contributions from gluon spin and 
orbital angular momenta are not still clear, so that various experimental
studies are in progress for clarifying the origin of the nucleon spin.

From the studies of the spin-1/2 nucleon, we found that our native 
models cannot be applied to hadron spin physics. Therefore, it
is important to test our knowledge on spin physics by other spin
observables. There exist new polarized structure functions for spin-one
hadrons \cite{fs83,hjm89,higher-spin}. There are also new fragmentation
functions \cite{spin-1-frag}, generalized parton distributions
\cite{spin-1-gpd}, and studies of target mass corrections
\cite{mass-corr} for spin-one hadrons.

Among the new structure functions, the leading-twist ones are
$b_1$ and $b_2$ which are related with each other by the Callan-Gross
type relation in the Bjorken scaling limit \cite{hjm89}.
It should be noted that $b_1$ vanishes if spin-1/2 constituents are
in the orbital S state, so that it is sensitive to dynamical aspects
of spin and orbital structure and possibly to non-nucleonic degrees
of freedom, for example, in the deuteron. In conventional approaches,
such tensor structure arises due to the D state admixture \cite{fs83,hjm89}, 
pions \cite{miller-b1}, and shadowing effects \cite{b1-shadowing,bj98}
in a nucleus. However, as it became obvious in the nucleon spin,
it is likely that high-energy tensor structure would not be simply
described by such conventional models. The new structure functions
could be important for probing unexplored dynamical aspects of
hadron spin.

The first measurement of the structure function $b_1$ was made by
the HERMES collaboration in 2005 \cite{hermes05}. The data indicated 
a finite distribution at $x<0.1$, which roughly agrees with
a double scattering contribution estimated in Ref. \cite{bj98}.
The data are also consistent with the quark-parton model sum rule
for $b_1$ \cite{b1-sum} although experimental errors are still large.
Positivity constraints are studied for $b_1$ 
in Ref. \cite{dmitrasinovic-96}.
In future, there are possibilities that the tensor structure
functions could be investigated in various facilities such as 
Thomas Jefferson National Accelerator Facility (JLab) and 
Japan Proton Accelerator Research Complex (J-PARC) \cite{j-parc}.

Although there are theoretical formalisms on the tensor
structure functions in polarized electron-hadron scattering
\cite{fs83,hjm89,higher-spin}, polarized hadron-hadron reactions,
such as at J-PARC, have not been well investigated for spin-one hadrons.
New polarized structure functions were found in a general formalism 
of polarized proton-deuteron Drell-Yan processes by imposing
time-reversal and parity invariances as well as Hermiticity
\cite{pd-drell-yan}. It should be noted that the tensor
distributions can be measured without polarizing the proton beam
\cite{pd-drell-yan}. 

It is important to obtain a reliable theoretical predication
for the tensor structure functions in order to compare with
experimental measurements.
In calculating structure functions for nuclei such as the deuteron,
a convolution model is often used for the hadron tensor $W^{\mu\nu}$,
which is given by the nucleonic tensor convoluted
with the spectral function of a nucleon in a nucleus \cite{sumemc}.
In order to extract each structure function from $W^{\mu\nu}$,
a corresponding projection operator needs to be multiplied \cite{projection}.
For the spin-$\frac{1}{2}$ nucleon, such projections have been already
studied \cite{projection,cps90}. It is the purpose of this article
to derive projection operators 
for the structure functions of a spin-one hadron. 
Our results should be useful for future theoretical studies
on spin-one hadrons.

This article consists of the following.
In Sec. \ref{spin-one}, a general framework of the structure functions
for a spin-one hadron is discussed. Then, the projection operators
are obtained in Sec. \ref{projection}, and results are summarized
in Sec. \ref{summary}.

\vspace{-0.1cm}
\section{Structure functions of spin-one hadrons}
\label{spin-one}
\vspace{-0.2cm}

A cross section for deep inelastic electron-hadron scattering
is described in terms of a hadron tensor $W_{\mu \nu}$.
The hadron tensor for a spin-$\frac{1}{2}$ target is expressed by
four structure functions $F_1$, $F_2$, $g_1$, and $g_2$
\cite{w_munu}:
\begin{gather}
W_{\mu\nu}^{\lambda_f \lambda_i} = -F_1 \hat{g}_{\mu \nu} 
+\frac{F_2}{M \nu} \hat{p}_\mu \hat{p}_\nu 
+ \frac{i g_1}{\nu}\epsilon_{\mu \nu \lambda \sigma} q^\lambda s^\sigma  
\nonumber \\
+\frac{i g_2}{M \nu ^2}\epsilon_{\mu \nu \lambda \sigma} 
       q^\lambda (p \cdot q s^\sigma - s \cdot q p^\sigma ),
\label{eqn:w-1/2}
\end{gather}
where $\epsilon_{\mu \nu \lambda \sigma}$ is an antisymmetric tensor
with the convention $\epsilon_{0123}=1$, $\nu$ is defined by
$\nu ={p \cdot q}/{M}$ with the hadron mass $M$, hadron momentum $p$,
and momentum transfer $q$, $Q^2$ is given by $Q^2=-q^2>0$,
and $s^\mu$ is the spin vector \cite{spin-vector}
which satisfies $s \cdot p=0$.
In Eq. (\ref{eqn:w-1/2}), we introduced notations:
\begin{equation}
\hat{g}_{\mu \nu} \equiv  g_{\mu \nu} -\frac{q_\mu q_\nu}{q^2}, \ \ 
\hat{a}_\mu \equiv a_\mu -\frac{a \cdot q}{q^2} q_\mu ,
\end{equation}
which ensure the current conservation 
$q^\mu W _{\mu \nu} = q^\nu W _{\mu \nu}=0$. 
The $a^\mu$ is a four vector, which is, for example, $p^\mu$ 
in Eq. (\ref{eqn:w-1/2}).
The coefficients of the unpolarized structure functions $F_1$ and $F_2$
are symmetric under $\mu \leftrightarrow \nu$ in $W_{\mu\nu}$,
while those of the polarized structure functions $g_1$ and $g_2$
are antisymmetric. 

In a spin-one hadron, there are four additional structure functions
$b_1$, $b_2$, $b_3$ and $b_4$ in the hadron tensor \cite{hjm89,w_munu}:
\begin{align}
& W_{\mu \nu}^{\lambda_f \lambda_i}
   = -F_1 \hat{g}_{\mu \nu} 
     +\frac{F_2}{M \nu} \hat{p}_\mu \hat{p}_\nu 
     -b_1 r_{\mu \nu} 
\notag \\
& \! \! \! \! 
     + \frac{1}{6} b_2 (s_{\mu \nu} +t_{\mu \nu} +u_{\mu \nu}) 
     + \frac{1}{2} b_3 (s_{\mu \nu} -u_{\mu \nu}) 
     + \frac{1}{2} b_4 (s_{\mu \nu} -t_{\mu \nu}) 
\notag \\ 
 & \! \! \! \! 
     + \frac{ig_1}{\nu}\epsilon_{\mu \nu \lambda \sigma} q^\lambda s^\sigma  
     +\frac{i g_2}{M \nu ^2}\epsilon_{\mu \nu \lambda \sigma} 
      q^\lambda (p \cdot q s^\sigma - s \cdot q p^\sigma ), 
\label{eqn:w-1}
\end{align}
where $r_{\mu \nu}$, $s_{\mu \nu}$, $t_{\mu \nu}$, and $u_{\mu \nu}$
are defined by
\begin{align}
r_{\mu \nu} = & \frac{1}{\nu ^2}
   \bigg [ q \cdot E ^* (\lambda_f) q \cdot E (\lambda_i) 
           - \frac{1}{3} \nu ^2  \kappa \bigg ]
   \hat{g}_{\mu \nu}, 
\notag \\ 
s_{\mu \nu} = & \frac{2}{\nu ^2} 
   \bigg [ q \cdot E ^* (\lambda_f) q \cdot E (\lambda_i) 
           - \frac{1}{3} \nu ^2  \kappa \bigg ]
\frac{\hat{p}_\mu \hat{p}_\nu}{M \nu}, \notag \\
t_{\mu \nu} = & \frac{1}{2 \nu ^2}
   \bigg [ q \cdot E ^* (\lambda_f) 
           \left\{ \hat{p}_\mu \hat E_\nu (\lambda_i) 
                 + \hat{p} _\nu \hat E_\mu (\lambda_i) \right\}
\notag \\  
 & + \left\{ \hat{p}_\mu \hat E_\nu^* (\lambda_f)  
           + \hat{p}_\nu \hat E_\mu^* (\lambda_f) \right\}  
     q \cdot E (\lambda_i) 
   - \frac{4 \nu}{3 M}  \hat{p}_\mu \hat{p}_\nu \bigg ] ,
\notag \\
u_{\mu \nu} = & \frac{M}{\nu} 
   \bigg [ \hat E_\mu^* (\lambda_f) \hat E_\nu (\lambda_i) 
          +\hat E_\nu^* (\lambda_f) \hat E_\mu (\lambda_i) 
\notag \\
& \ \ \ \ \ \  
   +\frac{2}{3}  \hat{g}_{\mu \nu}
   -\frac{2}{3 M^2} \hat{p}_\mu \hat{p}_\nu \bigg ],
\end{align}
where $\kappa= 1+{Q^2}/{\nu^2}$, and $s^\mu$ is 
the spin vector which satisfies $p \cdot s=0$.
The $E^\mu$ is the polarization vector of the spin-one hadron 
and it satisfies $p \cdot E =0 , E^* \cdot E =-1$.
The initial and final spin states are denoted by $\lambda_i$
and $\lambda_f$, respectively. 
Off-diagonal terms with $\lambda_f \ne \lambda_i$ need to be taken
into account in the general case to include higher-twist contributions
\cite{hjm89,higher-spin}.
The coefficients of $b_1$, $b_2$, $b_3$ and $b_4$ are symmetric under 
$\mu \leftrightarrow \nu$, and they vanish under the spin average. 
The functions $F_1$, $F_2$, $g_1$, and $g_2$ exist
in a spin-$\frac{1}{2}$ hadron as shown in Eq. (\ref{eqn:w-1/2}).
The $b_1$, $b_2$, $b_3$ and $b_4$ are new structure functions for a spin-one
hadron and they are associated with its tensor structure nature.

\vspace{-0.2cm}
\section{Projections to structure functions}
\label{projection}
\vspace{-0.2cm}

In the following calculations, we choose the frame
in which the target is at rest and the photon is moving
in the opposite direction to the $z$ axis. 
Then, the target-hadron and virtual-photon momenta are given
by $p^\mu =(M,0,0,0)$ and $q^\mu =(\nu,0,0,-| \vec q |)$,
respectively. However, results are Lorentz invariant, so that
they do not depend on the choice of the specific frame.

\vspace{-0.2cm}
\subsection{Spin-$\bf\frac{1}{2}$ hadrons}
\label{proj-spin-1/2} 
\vspace{-0.2cm}

Before discussing projections in a spin-one hadron, we first show
the spin-$\frac{1}{2}$ case, in which the structure functions
$F_1$, $F_2$, $g_1$, and $g_2$ exist. Projections
to these functions from $W_{\mu \nu}$ are discussed.
The hadron tensor $W_{\mu \nu}$ for a spin-$\frac{1}{2}$ hadron is given
in Eq. (\ref{eqn:w-1/2}).
Such projections were discussed in other articles \cite{projection,cps90};
however, they are explained in order to compare with the spin-one
projections in Sec. \ref{proj-spin-1}.

The spin vector is given by
$ \vec s_{\lambda_f \lambda_i}
  = N_{\lambda_f \lambda_i} u^\dagger _{\lambda_f}
    \vec \sigma u_{\lambda_i}$ \cite{hjm89}
where $\lambda_1$ and $\lambda_2$ are initial and final target spins,
respectively, along the $z$ axis, $\vec \sigma$ is the Pauli matrix,
$u_\lambda$ is the Pauli spinor, and 
$N_{\lambda_f \lambda_i}$ is a normalization factor to satisfy 
$ (\vec s_{\lambda_f \lambda_i})^* \cdot \vec s_{\lambda_f \lambda_i} =1$.
The spin four-vector is then given by
$
s^\mu_{\lambda_f \lambda_i} = (0, \vec s_{\lambda_f \lambda_i})
$
in the rest frame of the hadron.
Using 
$u_\uparrow   = \begin{pmatrix} 1 \\ 0 \end{pmatrix}$ and
$u_\downarrow = \begin{pmatrix} 0 \\ 1 \end{pmatrix}$, 
we have explicit expressions for the spin vectors
$s^\mu _{\uparrow \uparrow}$ and $s^\mu _{\uparrow \downarrow}$:
\begin{equation}
s^\mu _{\uparrow \uparrow}     = (0,0,0,1), \ \ 
s^\mu _{\uparrow \downarrow}   = \frac{1}{\sqrt{2}} (0,1,-i,0), \ \ 
s^\mu _{\downarrow \uparrow}   = ( s^\mu _{\uparrow \downarrow} )^* .
\end{equation}

In order to project out the four structure functions, four independent
combinations of momentum and spin need to be used. We choose a set: 
\begin{gather}
g^{\mu \nu}, \ \
\frac{p^\mu p^\nu}{M^2} , \ \ 
\frac{i}{M} \epsilon ^{\mu \nu \alpha \beta} 
          q_\alpha s_\beta ^{\uparrow \uparrow} 
          \delta_{\lambda_f \frac{1}{2}}
          \delta_{\lambda_i \frac{1}{2}},    
\notag \\
\frac{i}{M} \epsilon ^{\mu \nu \alpha \beta} 
          q_\alpha s_\beta ^{\uparrow \downarrow} 
          \delta_{\lambda_f -\frac{1}{2}}
          \delta_{\lambda_i \frac{1}{2}},           
\label{eqn:4-operators}
\end{gather}
for the projections from the hadron tensor $W_{\mu\nu}^{\lambda_f \lambda_i}$,
which depends on the initial and final spins through the spin vector $s^\mu$.
Here, $\delta_{\lambda \lambda^\prime}=1$ ($0$ )
for $\lambda=\lambda^\prime$ ($\lambda\ne\lambda^\prime$).
The first two terms in Eq. (\ref{eqn:4-operators}) are symmetric under
the exchange $\mu \leftrightarrow \nu$ and they project out 
the symmetric parts ($F_1$ and $F_2$) of $W_{\mu\nu}^{\lambda_f \lambda_i}$.
The latter two terms are antisymmetric and they project out 
the antisymmetric parts ($g_1$ and $g_2$).
Any other combinations such as
$p^\mu s^\nu _{\uparrow \uparrow }+p^\nu s^\mu _{\uparrow \uparrow }$
are not independent terms, so that there are only four independent
terms as shown in Eq. (\ref{eqn:4-operators}).
Using these terms, we can project 
out each structure function as follows:
\vspace{-0.2cm}
\begin{align}
\! \! \! \! \! 
F_1 & = -\frac{1}{2} \bigg ( g ^{\mu\nu} 
                - \frac{\kappa-1}{\kappa} \frac{p^\mu p^\nu}{M^2} \bigg ) 
         W_{\mu \nu}^{\lambda_f \lambda_i} ,
\notag \\
\! \! \! \! \! 
F_2 & = -\frac{x}{\kappa} 
           \bigg (g ^{\mu\nu} 
                - \frac{\kappa-1}{\kappa} \frac{3 p^\mu p^\nu}{M^2} \bigg ) 
         W_{\mu \nu}^{\lambda_f \lambda_i} ,
\notag \\
\! \! \! \! \! 
g_1 & = - \frac{i}{2 \kappa \nu}  
         \epsilon ^{\mu \nu \alpha \beta} q_\alpha 
   \left (  s_\beta^{\uparrow \uparrow} 
            \delta_{\lambda_f \frac{1}{2}}
            \delta_{\lambda_i \frac{1}{2}}   
          - s_\beta^{\uparrow \downarrow}
            \delta_{\lambda_f -\frac{1}{2}}
            \delta_{\lambda_i \frac{1}{2}}   
   \right ) W_{\mu \nu}^{\lambda_f \lambda_i} ,
\notag \\ 
\! \! \! \! \! 
g_2 & = \frac{i}{2 \kappa \nu}  
        \epsilon ^{\mu \nu \alpha \beta} q_\alpha
   \bigg ( s_\beta^{\uparrow \uparrow} 
            \delta_{\lambda_f \frac{1}{2}}
            \delta_{\lambda_i \frac{1}{2}}   
          +\frac{s_\beta^{\uparrow \downarrow}}{\kappa-1}
            \delta_{\lambda_f -\frac{1}{2}}
            \delta_{\lambda_i \frac{1}{2}}   
   \bigg ) W_{\mu \nu}^{\lambda_f \lambda_i} ,
\end{align}

$ \ \ \ $
\vspace{-0.5cm}

\noindent
where $x=Q^2/(2p\cdot q)$, and summations are taken over 
$\lambda_i$ and $\lambda_f$ in $g_1$ and $g_2$.

\vspace{-0.2cm}
\subsection{Spin-1 hadrons}
\label{proj-spin-1} 
\vspace{-0.2cm}

Next, we derive projections to the structure functions
for a spin-1 target. We use the following spherical unit vectors
as the target polarization vector \cite{edmond}:
\vspace{-0.2cm}
\begin{align}
  E^\mu (\lambda= \pm 1)   & = \frac{1}{\sqrt{2}}(0,\mp 1, -i,0),
\notag \\ 
  E^\mu (\lambda=0)   & = (0,0,0,1) .
\end{align}
For a spin-one hadron, the spin vector is expressed 
by the polarization vector $E^\mu$ as \cite{hjm89}
\vspace{-0.2cm}
\begin{equation}   
(s^{\lambda_f \lambda_i})^{\mu}
      = -\frac{i}{M} \epsilon ^{\mu \nu \alpha \beta} 
                E^*_\nu (\lambda_f) E_\alpha (\lambda_i) p_\beta ,
\end{equation}
where $M$ is the mass of the spin-1 hadron.
The initial and final polarization vectors are denoted by
$E^\mu (\lambda_i)$ and $E^\mu (\lambda_f)$, respectively,
with the spin states $\lambda_i$ and $\lambda_f$.
The spin vector $s^\mu$ is then given for
$s^\mu _{1 1}$, $s^\mu _{1 0}$, and $s^\mu _{0 1}$:
\vspace{-0.3cm}
\begin{equation}
s^\mu _{1 1}   = (0,0,0,1), \ \  
s^\mu _{1 0}   = \frac{1}{\sqrt{2}} (0,1,-i,0), \ \ 
s^\mu _{0 1}   = (s^\mu _{1 0})^* .
\end{equation}

$ \ \ \ $
\vspace{-0.5cm}

In projecting out the spin-1 structure functions, we may 
use only $\lambda=1$ and $\lambda=0$ terms because 
$\lambda=-1$ terms make the same contributions
as $\lambda=1$ ones. 
We follow a similar procedure for the projections
as the spin-$\frac{1}{2}$ case.
In the spin-one case, we have eight independent 
structure functions, so that eight combinations need to be taken.
We choose the following terms
\vspace{-0.3cm}
\begin{gather}
  g^{\mu\nu} \delta_{\lambda_f  1}
             \delta_{\lambda_i  1}   , \ \ 
  g^{\mu\nu} \delta_{\lambda_f  0}
             \delta_{\lambda_i  0}   , \ \ 
  g^{\mu\nu} \delta_{\lambda_f  1}
             \delta_{\lambda_i  0}   , 
\notag \\ 
  \frac{p^\mu p^\nu}{M^2} \delta_{\lambda_f  1}
                          \delta_{\lambda_i  1}   , \ \ 
  \frac{p^\mu p^\nu}{M^2} \delta_{\lambda_f  0}
                          \delta_{\lambda_i  0}   ,
\notag \\
  \frac{1}{M}
  [ p^\mu E ^\nu (\lambda =1) 
   +p^\nu E ^\mu (\lambda =1) ] \,  
   \delta_{\lambda_f  1}  \delta_{\lambda_i  0}   , 
\notag \\
  \frac{i}{M}
  \epsilon ^{\mu \nu \alpha \beta } 
  q_\alpha s_\beta ^{11}  \delta_{\lambda_f  1}
                          \delta_{\lambda_i  1} , \ \ 
  \frac{i}{M}
  \epsilon ^{\mu \nu \alpha \beta } 
  q_\alpha s_\beta ^{10}  \delta_{\lambda_f  0}
                          \delta_{\lambda_i  1} .
\end{gather}

$ \ \ \ $
\vspace{-0.6cm}

\noindent
The first six terms are associated with projections to
the structure functions $F_1$, $F_2$, $b_1$, $b_2$, $b_3$ and $b_4$,
and the last two terms are to $g_1$ and $g_2$. 
Off-diagonal terms with $\lambda_f \ne \lambda_i$ are needed
for including higher-twist effects.
Using these eight independent terms, we obtain the projections
from the hadron tensor in Eq. (\ref{eqn:w-1}) as
\vspace{-0.4cm}
\begin{widetext}
\vspace{-0.6cm}
\begin{align}
F_1 = & - \frac{1}{2} 
          \bigg ( g^{\mu \nu} 
                  - \frac{\kappa-1}{\kappa} \frac{p^\mu p^\nu}{M^2} \bigg )
   \frac{1}{3} 
   \delta_{\lambda_f \lambda_i}
   W_{\mu \nu} ^{\lambda_f \lambda_i} , \ \ \ \ \ \ \ \ \ \ \ \ 
F_2 =  -\frac{x}{\kappa} 
        \bigg (g ^{\mu\nu} 
               - \frac{\kappa-1}{\kappa} \frac{3 p^\mu p^\nu}{M^2} \bigg ) 
   \frac{1}{3} 
   \delta_{\lambda_f \lambda_i}
   W_{\mu \nu} ^{\lambda_f \lambda_i} , 
\notag \\
g_1 = & - \frac{i}{2 \kappa \nu} \epsilon ^{\mu \nu \alpha \beta} q_\alpha
         \bigg (  s_\beta ^{11} \delta_{\lambda_f  1}
                                \delta_{\lambda_i  1} 
                - s_\beta ^{10} \delta_{\lambda_f  0}
                                \delta_{\lambda_i  1} 
         \bigg ) W_{\mu \nu} ^{\lambda_f \lambda_i} , \ \ \ \ \ \ 
g_2 =    \frac{i}{2 \kappa \nu} \epsilon ^{\mu \nu \alpha \beta} q_\alpha
         \bigg (  s_\beta ^{11}  \delta_{\lambda_f  1}
                                 \delta_{\lambda_i  1} 
                + \frac{s_\beta ^{10}}{\kappa-1}
                                \delta_{\lambda_f  0}
                                \delta_{\lambda_i  1} 
         \bigg ) W_{\mu \nu} ^{\lambda_f \lambda_i} , \ \ \ 
\notag \\
b_1 = & \bigg [ -\frac{1}{2 \kappa}  g^{\mu \nu}
        \big ( \delta_{\lambda_f  0}
               \delta_{\lambda_i  0} 
              -\delta_{\lambda_f  1}
               \delta_{\lambda_i  0}  \big ) 
  + \frac{\kappa-1}{2 \kappa^2}  \frac{p^\mu p^\nu}{M^2} 
        \big ( \delta_{\lambda_f  0}
               \delta_{\lambda_i  0} 
              -\delta_{\lambda_f  1}
               \delta_{\lambda_i  1}  \big ) 
        \bigg ] W_{\mu \nu} ^{\lambda_f \lambda_i} , 
\notag \\
b_2= & \frac{x}{\kappa^2} \bigg [
        g^{\mu\nu} \bigg \{ - \delta_{\lambda_f  0}
                              \delta_{\lambda_i  0} 
                            - 2 (\kappa-1) \delta_{\lambda_f  1}
                                           \delta_{\lambda_i  1} 
                            + (2\kappa-1)  \delta_{\lambda_f  1}
                                           \delta_{\lambda_i  0} 
                   \bigg \}
       +\frac{3(\kappa-1)}{\kappa}  \frac{p^\mu p^\nu}{M^2} 
                 \big ( \delta_{\lambda_f  0}
                        \delta_{\lambda_i  0} 
                       -\delta_{\lambda_f  1}
                        \delta_{\lambda_i  1}  \big )
\notag \\
  &   \ \ \ \ \ \ \ \ \ \ \ \ \ \ \ 
       -\frac{4 (\kappa-1)}{\sqrt{\kappa} M}
          \big \{ p^\mu E^\nu (\lambda =1) + p^\nu E^\mu (\lambda =1) \big \}
                        \delta_{\lambda_f  1}
                        \delta_{\lambda_i  0}  \bigg ] 
        W_{\mu \nu} ^{\lambda_f \lambda_i},
\notag \\
b_3= & \frac{x}{3 \kappa^2} \bigg [
        g^{\mu\nu} \bigg \{ - \delta_{\lambda_f  0}
                              \delta_{\lambda_i  0} 
                            + \frac{2(2\kappa^2+2\kappa-1)}{\kappa-1} 
                              \delta_{\lambda_f  1}
                              \delta_{\lambda_i  1} 
                            - \frac{4\kappa^2+3\kappa-1}{\kappa-1}
                              \delta_{\lambda_f  1}
                              \delta_{\lambda_i  0} 
                   \bigg \}
       +\frac{3(\kappa-1)}{\kappa}  \frac{p^\mu p^\nu}{M^2} 
                 \big (  \delta_{\lambda_f  0}
                         \delta_{\lambda_i  0} 
                       - \delta_{\lambda_f  1}
                         \delta_{\lambda_i  1}  \big )
\notag \\
  &   \ \ \ \ \ \ \ \ \ \ \ \ \ \ \ 
       -\frac{4 (\kappa-1)}{\sqrt{\kappa} M}
          \big \{ p^\mu E^\nu (\lambda =1) + p^\nu E^\mu (\lambda =1) \big \}
                              \delta_{\lambda_f  1}
                              \delta_{\lambda_i  0}  \bigg ] 
        W_{\mu \nu} ^{\lambda_f \lambda_i},
\notag \\
b_4= & \frac{x}{3 \kappa^2} \bigg [
        g^{\mu\nu} \bigg \{ - \delta_{\lambda_f  0}
                              \delta_{\lambda_i  0} 
                            - \frac{2(\kappa^2+4\kappa+1)}{\kappa-1} 
                              \delta_{\lambda_f  1}
                              \delta_{\lambda_i  1} 
                            + \frac{2\kappa^2+9\kappa+1}{\kappa-1} 
                              \delta_{\lambda_f  1}
                              \delta_{\lambda_i  0} 
                   \bigg \}
       +\frac{3(\kappa-1)}{\kappa}  \frac{p^\mu p^\nu}{M^2} 
                 \big (  \delta_{\lambda_f  0}
                         \delta_{\lambda_i  0} 
                       - \delta_{\lambda_f  1}
                         \delta_{\lambda_i  1}  \big )
\notag \\
  &   \ \ \ \ \ \ \ \ \ \ \ \ \ \ \ 
       +\frac{4(2\kappa+1)}{\sqrt{\kappa} M}
          \big \{ p^\mu E^\nu (\lambda =1) + p^\nu E^\mu (\lambda =1) \big \}
                              \delta_{\lambda_f  1}
                              \delta_{\lambda_i  0}  \bigg ] 
        W_{\mu \nu} ^{\lambda_f \lambda_i} ,
\label{eqn:spin-1-proj}
\end{align}
\vspace{-0.1cm}
\end{widetext}
$ \ \ \ $
\vspace{-0.6cm}

\noindent
where summations are taken over $\lambda_i$ and $\lambda_f$.

It is useful to show the projections also in the Bjorken scaling limit,
$\nu, Q^2 \rightarrow \infty$ at finite $x=Q^2/(2 p\cdot q)$,
because the leading-twist structure functions such as $b_1$ and $b_2$
are first investigated experimentally.
In the scaling limit, we have the following relations
\begin{gather}
\ \ \vspace{-0.6cm}
   \lim_{\text{Bj}} g^{\mu\nu} W_{\mu \nu} ^{10} 
 = \lim_{\text{Bj}} g^{\mu\nu} W_{\mu \nu} ^{11} ,
\ \ \ \ 
   \lim_{\text{Bj}} \, (\kappa-1)
    \frac{p^\mu p^\nu}{M^2} W_{\mu \nu} ^{\lambda \lambda} = 0 ,
\nonumber \\
   \lim_{\text{Bj}} \,
    \frac{\kappa-1}{M} [ \, p^\mu E^\nu (\lambda=1)
                 +p^\nu E^\mu (\lambda=1) \, ] \, W_{\mu \nu} ^{10} = 0 ,
\end{gather}
$ \ \ \ $
\vspace{-0.45cm}

\noindent
by noting $\kappa \rightarrow 1$, $2x F_1 \rightarrow F_2$, and.
$2x b_1 \rightarrow b_2$.
Then, we obtain the following expressions for the leading-twist
structure functions:
\vspace{-0.2cm}
\begin{align}
F_1 = &  \frac{1}{2x} F_2 =  
   - \frac{1}{2} g^{\mu \nu} 
     \frac{1}{3} 
     \delta_{\lambda_f \lambda_i}
     W_{\mu \nu} ^{\lambda_f \lambda_i} ,
\notag \\
g_1 = & 
  - \frac{i}{2 \nu}
  \epsilon ^{\mu \nu \alpha \beta} 
  q_\alpha s_\beta ^{11} 
  \delta_{\lambda_f 1} \delta_{\lambda_i 1} 
  W_{\mu \nu} ^{\lambda_f \lambda_i} ,
\notag \\
b_1= & \frac{1}{2x} b_2 =  
   \frac{1}{2} g^{\mu \nu}
   \big ( \delta_{\lambda_f 1} \delta_{\lambda_i 1} 
         -\delta_{\lambda_f 0} \delta_{\lambda_i 0}  \big ) 
   W_{\mu \nu} ^{\lambda_f \lambda_i} ,
\end{align}
$ \ \ \ $
\vspace{-0.45cm}

\noindent
in the Bjorken scaling limit.

We have derived projections into the structure functions of
a spin-one hadron from its hadron tensor $W_{\mu\nu}$.
If a model is built for the hadron tensor, all the structure 
functions are derived without an approximation by using
the projections in Eq. (\ref{eqn:spin-1-proj}).

The spin structure of a spin-one hadron should be investigated
in future measurements. It should be possible to measure $b_1$
in the large-$x$ region at electron facilities such as JLab. 
The polarized proton-deuteron Drell-Yan processes should be
valuable in probing antiquark tensor polarizations
as unpolarized Drell-Yan measurements played a key role
in finding flavor dependence of antiquark distributions
in the nucleon \cite{b1-sum,flavor3}.
The measurement of the tensor polarized antiquark distributions
is possible without proton polarization \cite{pd-drell-yan}. 
There is a possibility to investigate such Drell-Yan processes
at J-PARC \cite{j-parc}. As the unexpected EMC spin measurement
led to many investigations on spin structure of spin-$\frac{1}{2}$
hadron, the tensor polarization studies could lead to a new dynamical
aspect of hadron spin physics.

\vspace{-0.5cm}
\section{Summary}
\label{summary}
\vspace{-0.3cm}

Spin structure of a spin-one hadron is interesting as a future
research topic because there exist new tensor structure functions which
do not appear in the spin-$\frac{1}{2}$ nucleon.
There are eight structure functions, 
$F_1$, $F_2$, $g_1$, $g_2$, $b_1$, $b_2$, $b_3$, and $b_4$,
in electron scattering from a spin-one hadron.
In this article, the projection operators have been derived
for the structure functions of a spin-one hadron by using
combination of its momentum, polarization, and spin vectors.
They are useful in theoretical calculations because the structure
functions need to be extracted from a calculated hadron tensor $W_{\mu\nu}$
in theoretical models.

\vspace{-0.5cm}
\begin{acknowledgements}
\vspace{-0.3cm}
One of the authors thanks H. Iwakiri and T.-H. Nagai for discussions.
\end{acknowledgements}

\vspace{-0.3cm}

\end{document}